\documentclass{emulateapj}
\slugcomment{To appear in \apj L}

\shorttitle{On the moat-penumbra relation}
\shortauthors{Vargas Dom\'inguez et al.}

\def\farcs{\hbox{$.\!\!^{\prime\prime}$}}

\begin{document}

 \title{On the moat-penumbra relation}

\author{S. Vargas Dom\'inguez\altaffilmark{1,2}, J.A. Bonet\altaffilmark{1}, V. Mart\'inez Pillet\altaffilmark{1}, Y. Katsukawa\altaffilmark{3}, Y. Kitakoshi\altaffilmark{3}, L. Rouppe van der Voort\altaffilmark{4,5}}

\altaffiltext{1}{Instituto de Astrof\'isica de Canarias, 38205 La Laguna, Tenerife, Spain; svargas@iac.es, jab@iac.es, vmp@iac.es}
\altaffiltext{2}{Department of Astrophysics, University of La Laguna, 38200, La Laguna, Tenerife, Spain.}
\altaffiltext{3}{National Astronomical Observatory of Japan, 2-21-1 Osawa, Mitaka, Tokyo 181-0033, Japan; yukio.katsukawa@nao.ac.jp, yasunori.kitakoshi@nao.ac.jp}
\altaffiltext{4}{Institute of Theoretical Astrophysics, University of Oslo, P.O. Box 1029 Blindern, N-0315 Oslo, Norway; v.d.v.l.rouppe@astro.uio.no}
\altaffiltext{5}{Center of Mathematics for Applications, University of Oslo, P.O. Box 1053 Blindern, N-0316 Oslo, Norway.}

\clearpage

\begin{abstract}
Proper motions in a sunspot group with a $\delta$-configuration and close to the
solar disc center have been studied by employing local correlation tracking
techniques. The analysis is based on more than one hour time series of G-band
images. Radial outflows with a mean speed of $0.67$ km s$^{-1}$ have been
detected around the spots, the well-known sunspots moats. However, these outflows are 
not found in those umbral core sides without penumbra. Moreover,
moat flows are only found in those sides of penumbrae located in the direction 
marked by the penumbral filaments. Penumbral sides perpendicular to them show no 
moat flow. These results strongly suggest a relation between the moat flow
and the well-known, filament aligned, Evershed flow. The standard picture
of a moat flow originated from a blocking of the upward propagation of
heat is commented in some detail.
\end{abstract}

\section{Introduction}

Sunspots, as complex magnetic structures embedded in a convective plasma,
show many active and changing features on multiple scales. Convective flows and
large-scale plasma circulation plays an important role in the dynamics and
evolution of solar active regions \citep[see, e.g.,][]{schrijver2000}
The granular convective pattern surrounding
sunspots is perturbed by the presence of magnetic elements known as Moving
Magnetic Features (MMFs), magnetic elements that move radially
outwards through an annular cell called the "moat" \citep[][for a recent view]{sheeley1972,harvey1973,hagenaar2005}
This moat flow also carries away
granules as inferred using local correlation techniques \citep[LCT,][]{bonet2005}.
It has been suggested that the moat could be a supergranule whose
center is occupied by a sunspot with a typical cell-scale of up to $10^4$ km
\citep{meyer1974}. The sunspot would act as a blocking agent for the 
upward propagation of heat from below. The excess temperature and
pressure generated in this way has been proposed as the origin of the 
moat flow \citep[see also][]{nye1988}. An interesting property of this supergranular cell
is that, in deeper layers, it would have a component directed towards the spot 
that could help stabilizing the magnetic structure.
While this viewpoint has been prevalent in sunspots physics, the real nature of the
moat flow is not well understood. Already \cite{vrabec1974} pointed out that in an irregular
sunspot, the moat was observed only in the sector with a well developed penumbra. 
The sunspot sector that had no penumbra, was an area where pores were being advected
(inflowing) {\it towards} the main spot and, thus, displaying no moat-like flow.

%%%%%%%%%%%%%%%%%%%%%%
%Fig. 1
\begin{figure}
\centering
\begin{tabular}{c}
\includegraphics[width=1.\linewidth]{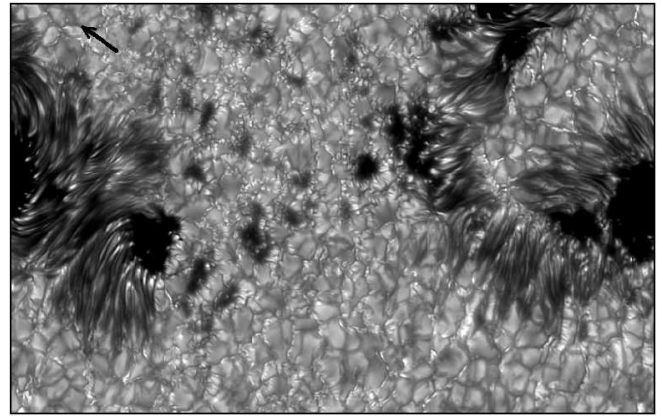} 
\end{tabular}
\vspace{-1mm}
\caption{G-band image of the field studied ($57\farcs8 \times 34\farcs4$),
reconstructed with the MOMFBD technique (see the text). The arrow in the upper
left corner points to the solar disk center.
\label{F:1}}
\end{figure}
%%%%%%%%%%%%%%%%%%%%%%
This old result already suggests a link between the presence of a penumbra and the 
moat flow, but no studies exist confirming such a relationship. 
It is interesting to point out that pores
are not seen to be surrounded by a moat (whereas mature sunspots 
always are). \cite{sobotka1999} studied (using LCT) six solar pores (including one with 
6 arcsec diameter) and found the surrounding motions to be dominated by mesogranular 
flows, as found elsewhere in the photosphere. No trace of a moat flow was observed.
It is clear that any possible heat blockage
by the pores magnetic body was unable to generate a moat flow.

The present paper studies the relation between moat flows and the 
existence of penumbrae.
It is based on an excellent $\sim$79 minutes time series of a
sunspot group with $\delta$-configuration. The $\delta$ configuration
is ideal for this study because it displays complex umbral regions
harboring fully developed penumbra on one side and none on the other.
The long duration of these series,
besides the high and stable quality throughout the entire period, substantially
improved after image reconstruction (see Fig.~\ref{F:1}), make of this material an
excellent data set to study the morphology and dynamical behaviour of sunspots
and their surroundings.

%%%%%%%%%%%%%%%%%%%%%%
%Fig. 2
\begin{figure*}
\centering
\begin{tabular}{ll}
\includegraphics[width=.8\linewidth]{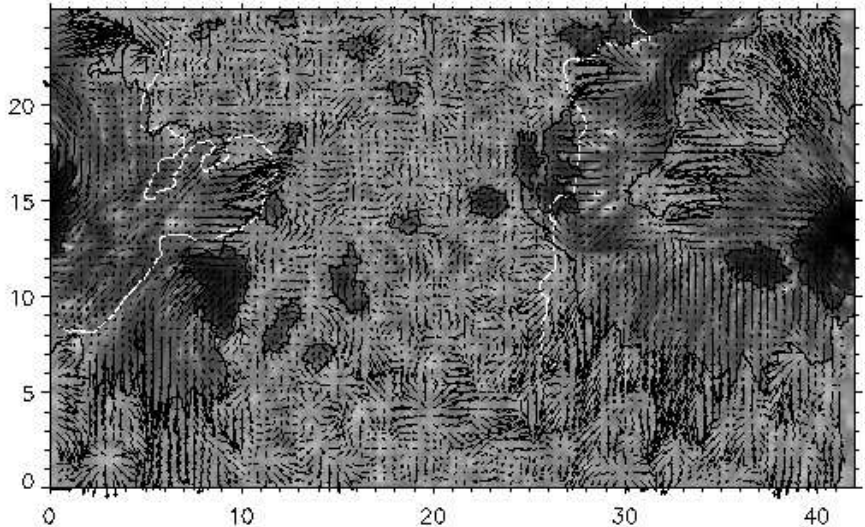} 
\end{tabular}
\vspace{-6mm}
\caption{Map of horizontal velocities in the entire FOV (71 min average). The
black contours outline the borders of umbrae, penumbrae and pores. The white
lines delineate the LOS neutral lines infered from a magnetogram. The
coordinates are expressed in Mm. The length of the black bar at coordinates
(0,0) corresponds to $0.4$ km s$^{-1}$. The background represents the average
image of the G-band series.
\label{F:2}}
%\vspace{1cm}
\end{figure*}
%%%%%%%%%%%%%%%%%%%%%%
\vspace{3mm}

\section{Observations and data processing}

The NOAA Active Region 10786 was observed on July 9, 2005 at the Swedish
1-meter Solar Telescope \citep[SST,][]{scharmer2003a} equipped with Adaptive
Optics (AO, Scharmer et al. 2003b), under collaboration in the International
Time Program. This complex region, corresponding to a $\delta$-configuration,
was placed at heliocentric position $\mu = 0.9$. A dichroic beamsplitter separates 
setups in the red and the blue. In the blue beam, two simultaneous time
sequences of high-resolution images were taken in wavelength bands at the
G-band ($430.5 \pm 0.54$ nm) and nearby continuum ($436.3 \pm 0.57$ nm,
G-cont). The sequences span for more than one hour, from 7:47 UT to 9:06 UT,
following the evolution of the sunspots. The images were acquired using Kodak
Megaplus 1.6 CCD cameras with a 10-bits dynamical range and 1536 $\times$ 1024
pixels. The pixel size was $0\farcs041$ square. Real-time corrections with the
AO and further post-processing techniques rendered up images near the
diffraction limit. Standard procedures for flatfielding and dark-current
subtraction were applied before restoration.

Post-processing for image restoration was performed by employing the
Multi-Object Multi-Frame Blind Deconvolution (MOMFBD) method \citep{noort2005}. The observational strategy to apply this
technique consisted in taking G-band images and simultaneous G-cont phase
diversity image-pairs \citep{gonsalves1982,paxman1992,lofdahl1994} by using an optical configuration with three channels. Sets of
about 18 images per channel (i.e. 3 $\times$ 18 images) were combined to
produce a pair of simultaneous G-band and G-cont restored images. Following
this procedure, two time series of reconstructed images were produced. Figure \ref{F:1} shows one of these images in the G-band. 

The restored images were de-rotated to compensate the diurnal field rotation
and rigidly aligned using cross-correlation. Additional image processing
consisted of destretching and p-mode filtering (threshold phase velocity $4$
km s$^{-1}$). The final product was two movies (G-band and G-cont) of 428
frames each, spanning over 71 min with a cadence of 10.0517 s, and covering a
field-of-view (FOV) of $57\farcs8 \times 34\farcs4$ (see movies at the web
site: \emph{http://www.iac.es/galeria/svargas/movies.html}).

\begin{figure}
%\centering
\hspace{-8mm}
\includegraphics[width=1.2\linewidth]{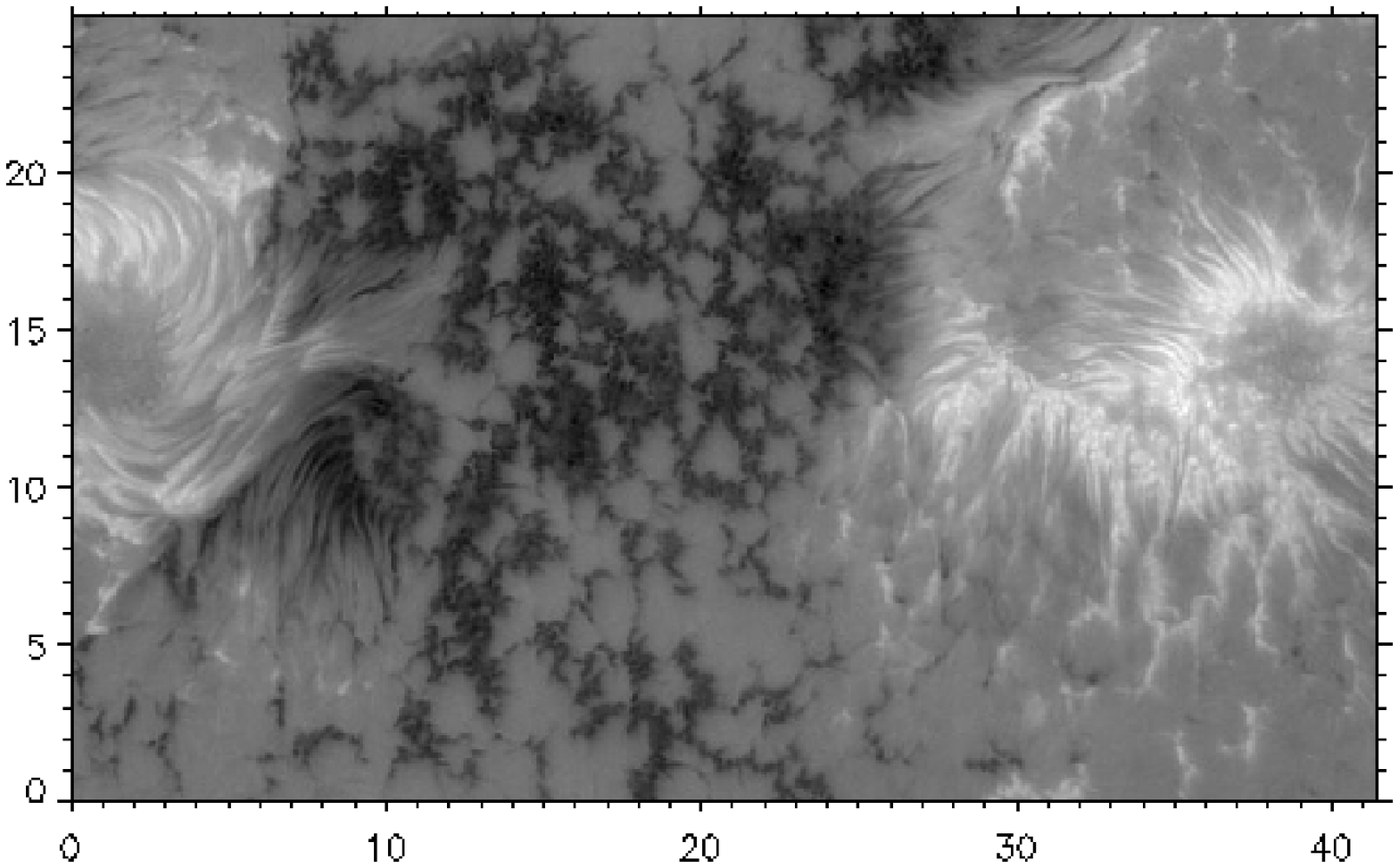}
\vspace{-8mm}
\caption{Corresponding magnetogram of the region under study obtained with the
SOUP filter. The $\delta$-spot neutral lines in Fig.~\ref{F:2} were obtained from this
frame. The coordinates are expressed in Mm.
\label{F:4}}
\end{figure}

In the red beam a tunable birefringent filter \citep[SOUP,][]{title1986}
working at FeI 6302\AA~in combination with liquid crystal retarders produced
longitudinal magnetograms. A beamsplitter in front of the SOUP filter splitted off
10\% of the light to obtain phase diversity image-pairs that combined with the
narrow-band images of the SOUP, allowed reconstructions MOMFBD to finally
provide magnetograms with $\sim 0\farcs25$  of resolution (see Fig.~\ref{F:4}).

%%%%%%%%%%%%%%%%%%%%%%
%Fig. 3
\begin{figure*}
\centering
\includegraphics[width=.8\linewidth]{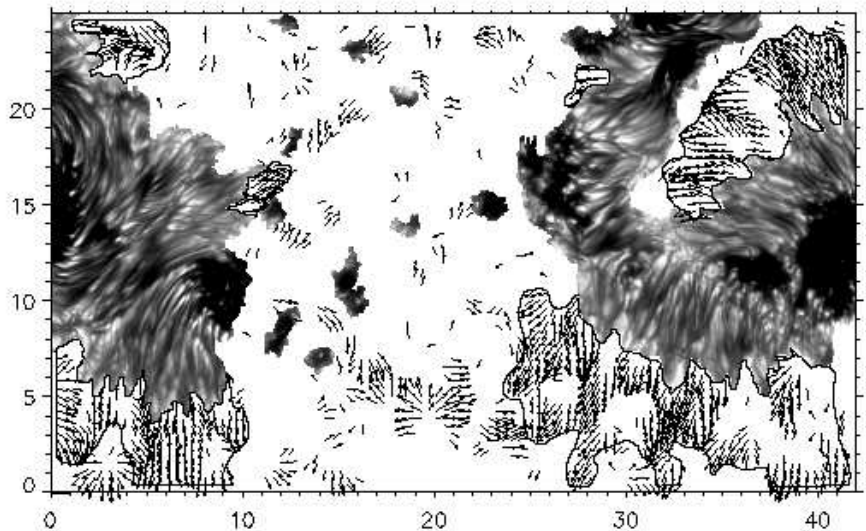}
\vspace{-6mm}
\caption{Map of the horizontal velocities with magnitude  $> 0.4$ km s$^{-1}$
(71 min average). The contrast within penumbrae has been enhanced (by removal
of a spatial running mean of the original image. Strong radial outflows (moats)
are evident surrounding all filamentary penumbrae. They are confined by the black contours.
These moats are found to be closely associated with the existence of penumbra.
Notice that moat region surrounding coordinates (1,5) corresponds to a
penumbra located to the left of the image and not visible in our FOV. The
coordinates are expressed in Mm. The length of the black bar at coordinates
(0,0) corresponds to $0.4$ km s$^{-1}$
\label{F:3}}
\end{figure*}
%%%%%%%%%%%%%%%%%%%%

%%%%%%%%%%%%%%%%%%%%%%
%Fig. 4
%%%%%%%%%%%%%%%%%%%%%

\section{Data analysis and results}
The G-band series has been used to study the proper motions of the structures
in the FOV by the local correlation tracking algorithm of \cite{november1988}, as implemented by \cite{molowny1994}. We have chosen a Gaussian
tracking window of FWHM $0\farcs78$ (half of the typical granular size)
suitable for tracking mainly granules. With this procedure, we obtain maps of
horizontal displacements or proper motions per time step (horizontal
velocities), which we average in time. Averages of horizontal velocities have
been performed over 5 min and 71 min intervals.

Figure \ref{F:2} shows the resulting flow map averaged over the whole series (71 min).
Moat flows are seen in the spot on the lower left of the FOV but more prominently
in the spot on the right side (upper and lower penumbral regions).
Several centers of diverging horizontal motions are present in the entire FOV.
Some surround sunspots (lower left and right sides of the figure) but the most
conspicuous ones are seen in the lesser magnetized area
(lower part of the FOV between the spots),
displaying greater velocities and more symmetrical shapes. 
These velocity structures, related to recurrent expansion and
splitting of granules, are commonly associated with mesogranulation \citep{roudier2003,roudier2004,bonet2005}. 

The white lines in Fig.~\ref{F:2} delineate the line-of-sight (LOS) neutral lines
inferred from a SOUP magnetogram (see Fig.~\ref{F:4}). The regions crossed by this line
correspond naturally to horizontal fields, similar to those found in penumbrae.
However, we caution that these horizontal fields can have different flows as
those that occur in normal penumbra (such as shear flows, as found by, \cite{deng2006} or the supersonic nozzles observed by \cite{pillet1994}.
Close inspection also shows that areas near neutral lines have a filamentary
appearance that is much less distinct than observed in normal penumbrae
extending radially from umbral cores.

The map in Fig.~\ref{F:3} shows only those velocities in the granulation field with
magnitude above $0.4$ km s$^{-1}$. Strong radial outflows (moats) are evident
surrounding the sunspots and $0.4$ km s$^{-1}$ appears to be a characteristic in
the velocity value to define the moats. The black contour in the figure outline
the area of the moats. It has been drawn by hand following visual criteria of
proximity to the sunspot and avoiding a few strong exploding events at the
lower part of the FOV that contaminate the large-scale moat flow.  Empty areas
in the moat region in the velocity maps are a result of the chosen value of the
velocity threshold and correspond to the interaction with exploding events
that occur inside the moat. 

From inspection of Fig.~\ref{F:3}, it becomes evident that the moats are closely
associated with the presence of a penumbra. The examples in the lower left, lower
right and upper right corners are quite revealing. The right pointed velocity
vectors on the lower left corner correspond to the moat flow of a penumbra that
is not visible in the FOV of this figure. Moats are missing in the sunspot
sides with no penumbral structure. This is readily seen in the right side of the 
umbral core at coordinates (9,10), but also in the left side of the one
at (25,18). None of the pores in the middle of the FOV is associated with any moat-like
flow. More importantly, moats are also absent in granulation regions
that are located side by side with a penumbra, but that are in the 
direction perpendicular to that marked by the penumbral filaments.
A clear example is the region around coordinates (9,6-7) in Fig.~\ref{F:3}. As a 'normal' 
(meaning here far from the neutral line) penumbra, it is a candidate to develop
a moat flow, but none is seen in this region. The moat flow only appears along the 
direction delineated by the penumbral filaments. Moats, thus, appear to
be a natural extension of the flow along the direction of penumbral filaments
and do not exist as outflows in the transverse direction.

The penumbral region in Fig.~\ref{F:3}, centered at coordinates (5,20), does not show clear 
evidences of a moat flow (although localized outflows are seen above and below it). 
This region corresponds to a strongly sheared neutral line commonly
seen in $\delta$ spots \citep{deng2006}. Figure \ref{F:4} shows a SOUP magnetogram obtained 
simultaneously with the G-band series. The spot centered on the left side of the FOV 
(white polarity) does not show a radially outward directed penumbra, but penumbral 
filaments that run parallel to the neutral line. This is a configuration 
commonly referred as sheared. This configuration is also inferred from the 
filaments observed in the G-band frames. We suspect that both conditions, the 
presence of a neutral line and the absence of radially oriented filaments, are
responsible for the absence of a clearly developed moat flow.
The nature of neutral lines with sheared configurations in $\delta$ spots
is not well understood. But we stress that if they correspond to relatively 
shallow structures, they would also block the upward propagation of heat and
should be able to generate a moat flow. Such a moat
flow is not observed in this sunspot. 

\section{Conclusions}
Time series of G-band and G-cont images of a sunspot group with
$\delta$-configuration (NOAA active region 10786), spanning over 71 min, were
observed and corrected for atmospheric and instrumental degradation. Proper
motions in the entire FOV have been measured in the G-band series by means of
local correlation tracking techniques. In this paper, we concentrate in the
results obtained in the identified moat flows (or lack thereof) of several
structures within the FOV. The main conclusions from our analysis
can be summarized as follows:

1) We have detected strong (mean speeds of $0.67$ km s${-1}$) outflows streaming
from penumbrae radially oriented from an umbral core, the so-called sunspot moats. 
Umbral cores sides with no penumbra {\em do not display moat flows}.

2) Furthermore, the moats are also absent on penumbral sides {\em perpendicular} to
the direction defined by the penumbral filaments. They are not found in 
directions transverse to them. A special case is sheared penumbral configurations
tangential to the umbral core. No moat flow is found there either.

These evidence is clearly suggestive of a link between the moat flow
and flows aligned with the penumbral filaments. The possible connection with the
Evershed flow is inescapable. Although it can be argued that a physical
relationship between these two flows is not firmly established in this work,
we believe the evidence is clear-cut and that a statistical analysis of 
a larger number of regions, as done here, can establish this connection 
more solidly. In particular, work is being done on analysing the moat flow 
configuration of round unipolar sunspots lacking penumbra in one of their sectors 
to find more evidence supporting points 1) and 2) above.

This result should be put in the context of the recent findings by \cite{sainz2005}; see also \cite{ravindra2006}. These authors find that the
penumbral filaments extend beyond the photometric sunspot boundary and cross
the region dominated by the moat flow. This region is also the location where the MMF
activity is normally found. Indeed some of these MMFs are seen by these authors
to start inside the penumbra. All these results point to an origin linking
the moat flow and MMF activity with the well-known Evershed flow. \cite{cabrera2006} also suggest that the Evershed clouds inside the penumbrae propagate to the moat becoming MMFs once they leave the sunspot.
It is indeed somewhat paradoxical that while the final fate of the Evershed
flow has remained unknown for decades, independent explanations and physical
scenarios have been proposed to generate the moat flow (that starts exactly
where the Evershed flow is seen to vanish). 

Similarly, the results obtained from local helioseismology near sunspots, in particular those
related to the f-mode \citep{gizon2000}, are also quite
relevant to the results presented here. The interpretation of these results
in terms of convective cells surrounding sunspots \citep[e.g.,][]{bovelet2003}
may be different if the Evershed flow turns out to be the major process
that injects mass into the moats surrounding sunspots.

We believe the hypothesis put
forward in this paper, the moat flow is the continuation outside the spots
of the Evershed flow, deserves adequate attention both from an observational and a theoretical point of view.

\acknowledgments{
The authors are grateful to Mats L\"ofdahl and Michiel Van Noort, for their
inputs about the restoration process. The Swedish 1-m Solar Telescope is
operated on the island of La Palma by the Institute of Solar Physics of the
Royal Swedish Academy of Sciences in the Spanish Observatorio del Roque de los
Muchachos of the Instituto de Astrof\'isica de Canarias. Partial support by the
Spanish Ministerio de Educaci\'on y Ciencia through project ESP2003-07735-C04 and 
financial support by the European Commission throught the SOLAIRE Network (MTRN-CT-2006-035484) 
are gratefuly \mbox{acknowledged}.
}


\begin{thebibliography}{}

\bibitem[Bonet et al.(2005)]{bonet2005} 
Bonet, J.A., M\'arquez, I., Muller, R., Sobotka, M., Roudier, Th., 2005, A\&A, 430, 1089

\bibitem[Bovelet \& Wiehr(2003)]{bovelet2003} 
Bovelet, B., \& Wiehr, E., 2003, A\&A 412, 249 

\bibitem[Cabrera Solana et al.(2006)]{cabrera2006} 
Cabrera Solana, D., Bellot Rubio, L.R., Beck, C., Del Toro Iniesta, J.C., ApJ 649, L41 

\bibitem[Deng et al.(2006)]{deng2006}
Deng, N., Xu, Y., Yang, G., Cao, W., Liu, C., Rimmele, T.R., Wang, H. \& Denker, C., 2006, ApJ 644, 1278

\bibitem[Gizon et al.(2000)]{gizon2000}
Gizon, L., Duvall, T.L., Jr., \& Larsen, R.M. 2000, J. Astrophys. Astron., 21, 339

\bibitem[Gonsalves(1982)]{gonsalves1982} 
Gonsalves, R. A., 1982, Opt. Eng. 21, 829

\bibitem[Hagenaar \& Shine(2005)]{hagenaar2005} 
Hagenaar, H.J., \& Shine, R.A., 2005, ApJ 635, 659 

\bibitem[Harvey \& Harvey(1973)]{harvey1973} 
Harvey, K., \& Harvey, J., 1973, Solar Phys. 28, 61

\bibitem[L\"ofdahl \& Scharmer(1994)]{lofdahl1994} 
L\"ofdahl, M.G., Scharmer, G.B., 1994, A\&A Suppl. Ser. 107, 243

\bibitem[Mart\'inez Pillet et al.(1994)]{pillet1994}
Mart\'inez Pillet, V., Lites, B. W., Skumanich, A., Degenhardt, D., 1994, ApJ, 425, 113 

\bibitem[Meyer et al.(1974)]{meyer1974}
Meyer, F., Schmidt, H.U., Weiss. N. O., Wilson, P.R., 1974, Proc. IAU Symp., 56, p.235

\bibitem[Molowny-Horas \& Yi(1994)]{molowny1994} 
 Molowny-Horas, R., \& Yi, Z., 1994, ITA (Oslo) Internal Rep. No. 31

\bibitem[November \& Simon(1988)]{november1988}
November, L.J., \& Simon, G.W., 1988, ApJ 333, 427

\bibitem[Nye et al.(1988)]{nye1988}
Nye, A., Bruning, D., Labonte, B. J., 1988, Solar Phys. 115, 251

\bibitem[Paxman et al.(1992)]{paxman1992} 
Paxman, R.G., Schulz, T.J., Fienup, J.R., 1992, J.Opt.Soc.Am. A9, 7, 1072

\bibitem[Ravindra(2006)]{ravindra2006}
Ravindra, B., 2006, Solar Phys. 236, 297

\bibitem[ Roudier \& Muller(2004)]{roudier2004} 
 Roudier, Th., \& Muller, R., 2004, A\&A, 419, 757
		
\bibitem[Roudier et al.(2003)]{roudier2003} 
Roudier, Th., Ligni\`eres, F., Rieutord, M., Brandt, P.N., \& Malherbe, J.M. 2003, A\&A, 409, 299	

\bibitem[Sainz Dalda \& Mart\'inez Pillet(2005)]{sainz2005}
Sainz Dalda, A., Mart\'inez Pillet, V., 2005, ApJ, 662, 1176

\bibitem[Scharmer et al.(2003a)]{scharmer2003a}
Scharmer G. B., Bjelksj\"o K., Korhonen, T.K., Lindberg, B. \& Petterson, B., 2003a, Proc. SPIE. 4853, 341

\bibitem[Scharmer et al.(2003b)]{scharmer2003b}
Scharmer G. B., Dettori, P. M., L\"ofdahl, M. G. \& Shand, M., 2003b, Proc. SPIE, 4853, 370

\bibitem[Schrijver \& Zwann(2000)]{schrijver2000} 
Schrijver, C. J., \& Zwann, C., 2000, Solar and stellar magnetic activity. Cambridge U Press

\bibitem[Sheeley(1972)]{sheeley1972} 
Sheeley, N.R., 1972, Solar Phys. 25, 98

\bibitem[Sobotka et al.(1999)]{sobotka1999}
Sobotka, M., V\'azquez, M., Bonet, J.A., Hanslmeier, A., Hirzberger, J., 1999, ApJ, 511, 436

\bibitem[Title et al(1986)]{title1986}
Title, A. M., Tarbell, T., Simon, G., and the SOUP team, 1986, Adv. Space Res., 6, 253

\bibitem[Van Noort et al.(2005)]{noort2005}
Van Noort. M., Rouppe van der Voort, L., L\"ofdahl, M. G.,  2005, Sol. Phys., 228, 191

\bibitem[Vrabec(1974)]{vrabec1974}
Vrabec, D., 1974, Proc. IAC Symp., 56, p.201


\end{thebibliography}
\end{document}